\documentclass[journal]{IEEEtran}

\usepackage{amsmath,amsfonts}
\usepackage{algorithmic}
\usepackage{algorithm}
\usepackage{array}
\usepackage{textcomp}
\usepackage{url}
\usepackage{verbatim}
\usepackage{graphicx}
\usepackage{cite}
\usepackage{siunitx}
\usepackage{subcaption}
\usepackage{comment}
\usepackage[acronym]{glossaries}
\usepackage{tabularx}
\usepackage{booktabs}
\usepackage[table, dvipsnames]{xcolor}
\usepackage{framed}

\definecolor{verylightgray}{RGB}{240, 240, 240}       
\newacronym{ADC}{ADC}{Analog-to-Digital Converter}
\newacronym{AI}{AI}{Artificial Intelligence}
\newacronym{ARIMA}{ARIMA}{AutoRegressive Integrated Moving Average}
\newacronym{API}{API}{Application Programming Interface}

\newacronym{CP}{CP}{Control Plane}
\newacronym{CPU}{CPU}{Central Processing Unit}
\newacronym{CU}{CU}{Centralized Unit}

\newacronym{DAC}{DAC}{Digital-to-Analog Converter}
\newacronym{dApp}{dApp}{Distributed Application}
\newacronym{DU}{DU}{Distributed Unit}

\newacronym{eBPF}{eBPF}{extended Berkeley Packet Filter}

\newacronym{FEC}{FEC}{Forward Error Correction}
\newacronym{FFT}{FFT}{Fast Fourier Transform}
\newacronym{FIFO}{FIFO}{First-In, First-Out}

\newacronym{gNB}{gNB}{Next Generation NodeB}

\newacronym{IPC}{IPC}{Instructions Per Cycle}

\newacronym{near-RTRIC}{near-RT RIC}{near-RealTime RAN Intelligence Controller}
\newacronym{non-RTRIC}{non-RT RIC}{non-RealTime RAN Intelligence Controller}
\newacronym{NUMA}{NUMA}{Non-Uniform Memory Access}

\newacronym{MAC}{MAC}{Medium Access Control}
\newacronym{MDP}{MDP}{Markov Decision Process}
\newacronym{MIMO}{MIMO}{Multiple Input Multiple Output}
\newacronym{ML}{ML}{Machine Learning}
\newacronym{MPKI}{MPKI}{Misses Per Kilo-Instruction}

\newacronym{OAI}{OAI}{OpenAirInterface}
\newacronym{OCU}{O-CU}{O-RAN CU}
\newacronym{ODU}{O-DU}{O-RAN DU}
\newacronym{ORAN}{O-RAN}{Open RAN}
\newacronym{ORU}{O-RU}{O-RAN RU}
\newacronym{OS}{OS}{Operating System}

\newacronym{PHY}{PHY}{Physical}

\newacronym{QoS}{QoS}{Quality of Service}

\newacronym{RAN}{RAN}{Radio Access Network}
\newacronym{RAM}{RAM}{Random Access Memory}
\newacronym{rApp}{rApp}{RAN Application}
\newacronym{RF}{RF}{Radio Frequency}
\newacronym{RIC}{RIC}{RAN Intelligent Controller}
\newacronym{RL}{RL}{Reinforcement Learning}
\newacronym{RR}{RR}{Round Robin}
\newacronym{RU}{RU}{Radio Unit}
\newacronym{RT-RIC}{RT-RIC}{Real-Time RIC}

\newacronym{SDR}{SDR}{Software-Defined Radio}
\newacronym{SMO}{SMO}{Service Management and Orchestration}
\newacronym{SMT}{SMT}{Simultaneous Multithreading}

\newacronym{TCP}{TCP}{Transmission Control Protocol}
\newacronym{TS}{TS}{Time Sharing}
\newacronym{TTI}{TTI}{Transmission Time Interval}

\newacronym{UE}{UE}{User Equipment}
\newacronym{UP}{UP}{User Plane}
\newacronym{UPF}{UPF}{User Plane Function}

\newacronym{vBS}{vBS}{virtualized Base Station}
\newacronym{vRAN}{vRAN}{Virtualized RAN}

\newacronym{xApp}{xApp}{eXtended Application}

\newacronym{ZMQ}{ZMQ}{ZeroMQ}

\begin{document}

\title{Energy-Aware CPU Orchestration in O-RAN: A dApp-Driven Lightweight Approach}

\author{
    \IEEEauthorblockN{
        Francisco Crespo\IEEEauthorrefmark{1},
        Javier Villegas\IEEEauthorrefmark{1},
        Carlos Baena\IEEEauthorrefmark{1},
        Eduardo Baena\IEEEauthorrefmark{2},
        Sergio Fortes\IEEEauthorrefmark{1},
        Raquel Barco\IEEEauthorrefmark{1}
    }

    \IEEEauthorblockA{
        \IEEEauthorrefmark{1}Telecommunication Research Institute (TELMA), Universidad de Málaga, E.T.S. Ingeniería de Telecomunicación, Bulevar Louis Pasteur 35, 29010, Málaga, Spain\\
        \IEEEauthorrefmark{2}Institute for the Wireless Internet of Things, Northeastern University, Boston, MA, USA
    }
    
    \thanks{Corresponding author: Sergio Fortes (e-mail: sfr@ic.uma.es).}
    \thanks{This work has been partially funded by: Ministerio de Asuntos Económicos y Transformación Digital and European Union - NextGenerationEU within the framework ``Recuperación, Transformación y Resiliencia y el Mecanismo de Recuperación y Resiliencia'' under the project MAORI. This work has been also supported by Ministerio de Ciencia y Tecnología through grant FPU21/04472.}
}

\maketitle

\begin{abstract}
The transition toward softwarized Radio Access Networks (RANs), driven by the Open RAN (O-RAN) paradigm, enables flexible, vendor-neutral deployments through disaggregation and virtualization of base station functions. However, this shift introduces new challenges in managing CPU resources efficiently under strict real-time constraints. In particular, the interplay between latency-sensitive RAN workloads and general-purpose Operating System (OS) schedulers often leads to suboptimal performance and unnecessary energy consumption. This work proposes a lightweight, programmable distributed application (dApp) deployed at the Distributed Unit (DU) level to dynamically orchestrate CPU usage. The dApp operates in closed loop with the OS, leveraging thread-level telemetry like context switches, Instructions Per Cycle (IPC), and cache metrics, to adapt CPU thread affinity, core isolation, and frequency scaling in real time. Unlike existing solutions, it requires no access to proprietary RAN software, hardware-specific features, or kernel modifications. Fully compliant with the O-RAN architecture and agnostic to the underlying RAN stack, the proposed solution introduces negligible overhead while improving energy efficiency and CPU utilization. Experimental results using a commercial-grade srsRAN deployment demonstrate consistent power savings without compromising real-time processing performance, highlighting the potential of low-latency dApps for fine-grained resource control in next-generation networks.
\end{abstract}


\begin{picture}(0,0)(0,-485)
  \put(20,0){
    \setlength{\fboxsep}{5pt} 
    \fbox{
      \begin{minipage}{0.8\textwidth}
        \footnotesize
        \centering
        This work has been submitted to the IEEE for possible publication.\\
        Copyright may be transferred without notice, after which this version may no longer be accessible.
      \end{minipage}
    }
  }
\end{picture}

\section{Introduction}

\IEEEPARstart{T}{he} increasing demand for mobile data, along with the proliferation of emerging services, is pushing operators to continuously evolve their \gls{RAN} infrastructures. In this context, the disaggregation of \gls{RAN} components, central to the \gls{ORAN} paradigm, has emerged as a key strategy to promote vendor-neutral deployments, accelerate innovation, and reduce operational costs by leveraging standardized interfaces and cloud-native implementations~\cite{oran_alliance}.

While the softwarization of \gls{RAN} introduces substantial flexibility and programmability, it also exposes the system to new performance and orchestration challenges. Among them, the dynamic allocation of \gls{CPU} resources becomes critical for maintaining \gls{QoS} guarantees, particularly for time-sensitive \gls{RAN} tasks. These tasks, such as \gls{FEC} or \gls{FFT} operations, must meet strict deadlines at the physical layer while sharing hardware with other network functions and background processes~\cite{oRan, Perfprofiling}.

To manage this orchestration complexity, the O-RAN architecture introduces a layered control framework centered on \glspl{RIC}. The \gls{non-RTRIC} provides long-term policy optimization, while the \gls{near-RTRIC} supports control loops operating on the order of tens to hundreds of milliseconds via programmable applications such as \glspl{xApp} and \glspl{rApp}. However, these controllers operate above the latency threshold required for fine-grained coordination with per-thread CPU scheduling or execution-level decisions within the \gls{DU}.

To bridge this latency and visibility gap, recent efforts have proposed a new class of control logic known as distributed applications (\glspl{dApp}), which execute directly at the \gls{DU} level. dApps extend the O-RAN control architecture by enabling sub-10 millisecond inference and control, with access to rich runtime data such as user-plane metrics, I/Q samples, and scheduling queues~\cite{dappsdef,oran_whitepaper,arxiv_dApp}. In contrast to xApps and rApps, which operate in centralized controllers, dApps execute natively within the target node, enabling fast, closed-loop reactions to dynamic system behavior without incurring additional signaling latency.

In this work, we propose a lightweight, \gls{ORAN}-compliant \gls{dApp} designed to perform fine-grained \gls{CPU} management directly within the \gls{DU}. Specifically, the proposed \gls{dApp} dynamically orchestrates \gls{OS}-level mechanisms such as CPU affinity, thread isolation, and frequency scaling for softwarized \gls{RAN} workloads running on commodity hardware. It leverages low-level telemetry, collected through standard Linux system tools, including context switches, instructions per cycle (IPC), cache miss metrics (MPKI), and power consumption data. This enables adaptive control based on traffic and workload dynamics.

By closing the loop between \gls{RAN}-level orchestration and \gls{OS}-level \gls{CPU} management, our \gls{dApp} achieves energy-aware scheduling without modifying the underlying \gls{RAN} software stack or relying on external \gls{RIC} components. Moreover, by operating entirely within the timing budget of \gls{DU}-level threads, it provides a practical and responsive control solution for performance-critical environments.

\textbf{The main contributions of this paper are:}
\begin{itemize}
    \item We propose an \gls{ORAN}-compliant architecture that integrates fine-grained \gls{CPU} control within the \glspl{DU} via a lightweight, containerized \gls{dApp}, deployable on commercial \gls{RAN} software stacks.
    
    \item We develop a measurement framework using \texttt{perf} to capture thread-level execution metrics such as context switches, \gls{IPC}, and \gls{MPKI}, and correlate them with power consumption and RAN performance indicators.

    \item We implement a dynamic CPU frequency control strategy within the \gls{dApp} and evaluate its effectiveness under varying traffic and load conditions. We also conduct detailed profiling to assess the impact of thread migration and core affinity.

    \item We experimentally validate the proposed approach using srsRAN, an open-source \gls{RAN} software stack, demonstrating consistent energy savings without degradation of real-time processing performance.
\end{itemize}

\textbf{Paper organization:} Section~\ref{sec:related} reviews prior work on \gls{RAN} function profiling and energy-aware orchestration in the context of \gls{ORAN}. Section~\ref{sec:problem} formulates the CPU scheduling problem and presents our measurement methodology. Section~\ref{sec:architecture} details the system architecture and \gls{dApp} design. Section~\ref{sec:evaluation} presents experimental validation and analysis. Finally, Section~\ref{sec:conclusion} concludes the paper and outlines future work.


\section{Related Work}\label{sec:related}

The softwarization of \gls{RAN} functions introduces significant challenges in resource allocation and energy efficiency, especially under strict real-time constraints. These challenges span both system-level orchestration and low-level execution behavior, and have been partially addressed in prior work from different perspectives.

At the system level, studies such as \cite{Maria2023} and \cite{Sep2023} evaluate the impact of radio configuration and computational constraints in softwarized environments using commercial platforms like Amarisoft~\cite{Amarisoft}. Their results show that increasing bandwidth or deploying \gls{MIMO} does not always yield performance gains when \gls{CPU} availability is limited. Moreover, the authors in \cite{vRANBaena} analyze how constrained \gls{RAN} computational resources affect service-level Quality of Experience (QoE). Additional modeling of CPU load under multi-user conditions is presented in \cite{feb2024}, which proposes regression techniques to estimate RAN performance degradation. Further, \cite{jmila2017estimating} proposes a \gls{ML}-based method to estimate virtualized network function resource demands.

Beyond high-level system behavior, low-level profiling of softwarized base stations has emerged as a powerful tool to understand computational bottlenecks. In \cite{Perfprofiling}, the authors use \textit{perf} (Performance Counters for Linux) to analyze the behavior of a 5G stack, highlighting the processing cost of \gls{PHY}-layer tasks and the limited overhead introduced by the CU/DU split. Similarly, \cite{AIRIC} leverages \texttt{perf} metrics, such as context switches and CPU migrations, to detect performance degradation caused by co-located processes (i.e., noisy neighbors), using neural network classifiers.

However, most of these profiling efforts operate post-factum or at coarse time resolutions. They lack real-time actuation and are not directly integrated into orchestration mechanisms. In contrast, our work focuses on actionable, fine-grained telemetry that can drive real-time CPU scheduling decisions without prior instrumentation.

Energy optimization via OS-level control policies has also been explored in recent work. Several RAN implementations such as Amarisoft~\cite{amarisoft_linux_setup}, srsRAN~\cite{srsranDocs}, and \gls{OAI}~\cite{KALTENBERGER2020107284} recommend using the Linux \textit{performance} governor to avoid deadline violations, but this leads to maximum CPU frequencies regardless of actual load, increasing energy consumption unnecessarily. To mitigate this, RENC~\cite{renc_nsdi} introduces slack-aware frequency scaling using \gls{eBPF}, avoiding deep C-states. However, it requires access to internal RAN stack metrics, which is infeasible in black-box deployments.

Unlike RENC, the approach proposed in this work does not rely on modifying the kernel or instrumenting RAN threads. Instead, it infers scheduling inefficiencies and CPU stress from observable metrics such as \gls{IPC}, \gls{MPKI}, and context switches. Moreover, it jointly addresses frequency scaling, thread–core affinity, and dynamic core isolation, dimensions that are typically considered in isolation in prior work.

Recent proposals like \cite{urumkar23} and \cite{urumkar2_2023} address CPU scheduling and fault tolerance in \gls{ORAN}, but focus on heuristic strategies and control-plane orchestration, without engaging with low-level runtime behavior. Similarly, \cite{pamuklu21} and \cite{liang2024} propose energy-aware function placement strategies, but not at the OS-level scheduling granularity targeted in this work.

Centralized orchestration solutions such as AIRIC~\cite{AIRIC}, which operate at the \gls{SMO} level and aggregate telemetry across multiple nodes, introduce additional abstraction layers and control latency. In contrast, our proposal is situated at the execution layer and enables near-real-time actuation over local CPU scheduling decisions. To the best of the authors’ knowledge, this is the first work to integrate low-level CPU telemetry, dynamic affinity control, and real-time frequency tuning within an \gls{ORAN}-compliant \gls{dApp} deployed directly at the \gls{ODU}.

\section{Problem Formulation}
\label{sec:problem}

This section formulates the core technical challenge addressed in this work: how to minimize software-based RAN \gls{CPU} energy consumption without violating the strict timing constraints imposed by \gls{RAN} workloads (i.e., \gls{TTI} deadlines and throughput requirements).

To formally represent the energy consumption of the system, we adopt the widely-used statistical power consumption model presented in~\cite{PowerConsumptionModel}. This model is given by:
\begin{equation}
P_f = P_s + kf^2,
\label{eq:powerEq}
\end{equation}
where $P_f$ is the power consumption of the \gls{CPU} at frequency $f$, $P_s$ is the static power term representing the baseline power required by the cores to operate, and $kf^2$ is the dynamic power, with $k$ being a hardware-dependent constant that modulates how consumption scales with frequency. According to this model, aggressively running the system at peak frequency quickly escalates the dynamic power term, leading to high energy consumption. Conversely, operating at lower frequencies saves power but increases the risk of missing \gls{RAN} task deadlines, as both \gls{PHY} and \gls{MAC} layers must complete their processing within each \gls{TTI}.

\begin{figure}[t]
\centering
\begin{subfigure}[t]{\columnwidth}
\includegraphics[width=\linewidth]{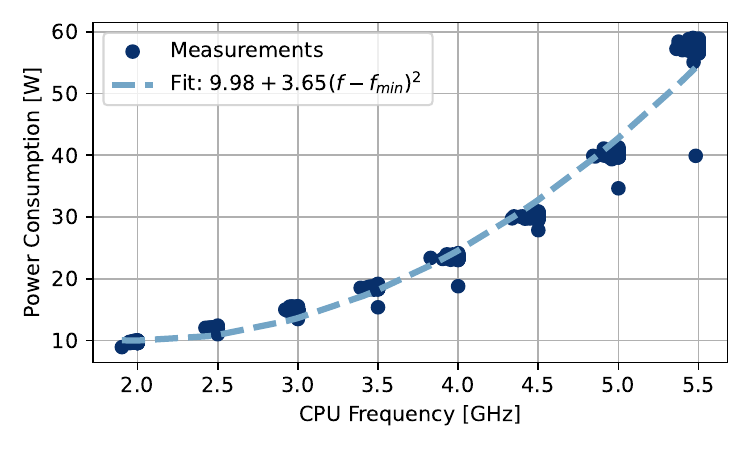}
\caption{Relation between \gls{CPU} frequency and power consumption.}
\label{fig:power_vs_freq}
\end{subfigure}

\vspace{0.5em}
\begin{subfigure}[t]{\columnwidth}
    \includegraphics[width=\linewidth]{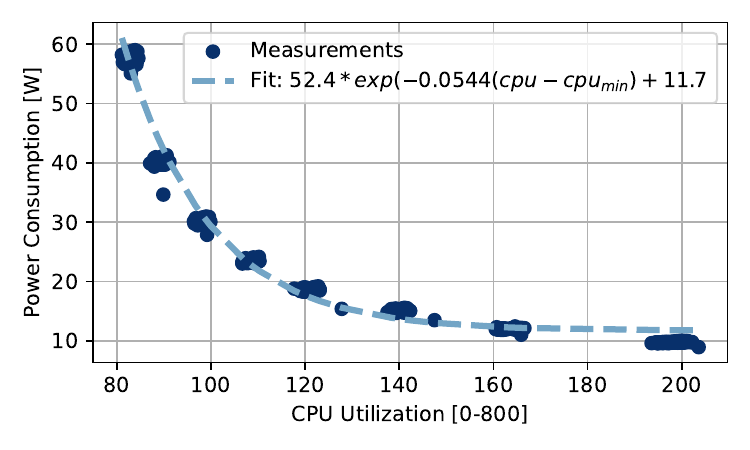}
    \caption{Relation between \gls{CPU} utilization and power consumption.}
    \label{fig:power_vs_util}
\end{subfigure}
\caption{Relation between power consumption against \gls{CPU} frequency and utilization.}
\label{fig:power_model}
\end{figure}

To validate the applicability of this power model in our scenario, we experimentally characterized the relationship between \gls{CPU} frequency, utilization, and power consumption on a server running srsRAN, as depicted in Figure~\ref{fig:power_model}. Specifically, Figure~\ref{fig:power_vs_freq} confirms the quadratic relation between frequency and power consumption predicted by the dynamic term of the model, while Figure~\ref{fig:power_vs_util} shows the exponential growth of power consumption with increased \gls{CPU} utilization. These observations emphasize the non-linear relationship between utilization, frequency, and energy efficiency, reinforcing the need for dynamic and intelligent resource management.

Considering common approaches to minimize the power consumption of multi-core \glspl{CPU}, some methods focus on reducing the static power $P_s$ through core deactivation, power gating, or using deeper C-states. Others aim to lower the dynamic power term by applying Dynamic Voltage and Frequency Scaling (DVFS) across all cores, leveraging execution slack and limiting throughput at the \gls{MAC} layer to make such transitions profitable~\cite{renc_nsdi}. However, the time required to transition in and out of deeper C-states or between different frequencies imposes a strict constraint in real-time systems:
\begin{equation}
\max(RT, WT) < \text{TTI},
\label{eq:timeState}
\end{equation}
where $RT$ (Residency Time) is the minimum time required for a CPU state to justify the extra transition energy overhead, and $WT$ (Wakeup Time) represents the latency to resume from a sleep state. Given typical \gls{TTI} durations ranging from $1\,ms$ down to $62.5\,\mu s$, frequent transitions are severely constrained, making fine-grained DVFS and state transitions impractical.

\begin{figure}[t]
    \centering
    \includegraphics[width=\linewidth]{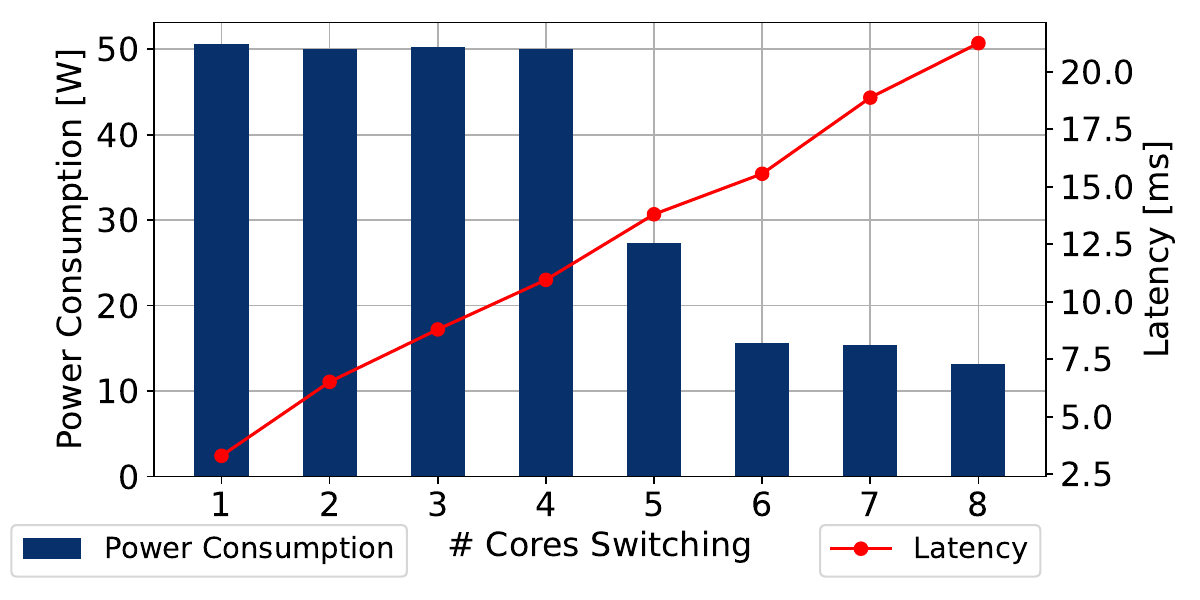}
    \caption{Power consumed and response latency at maximum switching speed.}
    \label{fig:freq_switch}
\end{figure}

To quantify this further, we experimentally analyzed the energy cost and latency of maximum-frequency switching rates, as illustrated in Figure~\ref{fig:freq_switch}. These results demonstrate the significant latency and energy penalty incurred when changing frequencies frequently, reinforcing the conclusion of previous works~\cite{renc_nsdi,FrequencyAffinity} that constant frequency switching is computationally expensive and energy inefficient. Thus, a more appropriate approach is frequency affinity: maintaining constant voltage and frequency as long as possible and limiting frequency transitions only to essential cases. Following the strategy described in~\cite{FrequencyAffinity}, thread-level scheduling informed by memory access patterns can effectively cluster threads with similar characteristics, facilitating energy-efficient frequency selection and affinity settings.

\begin{figure*}[h]
\centering
\includegraphics[width=0.75\textwidth,clip=True,trim={36pt 0pt 0pt 0pt}]{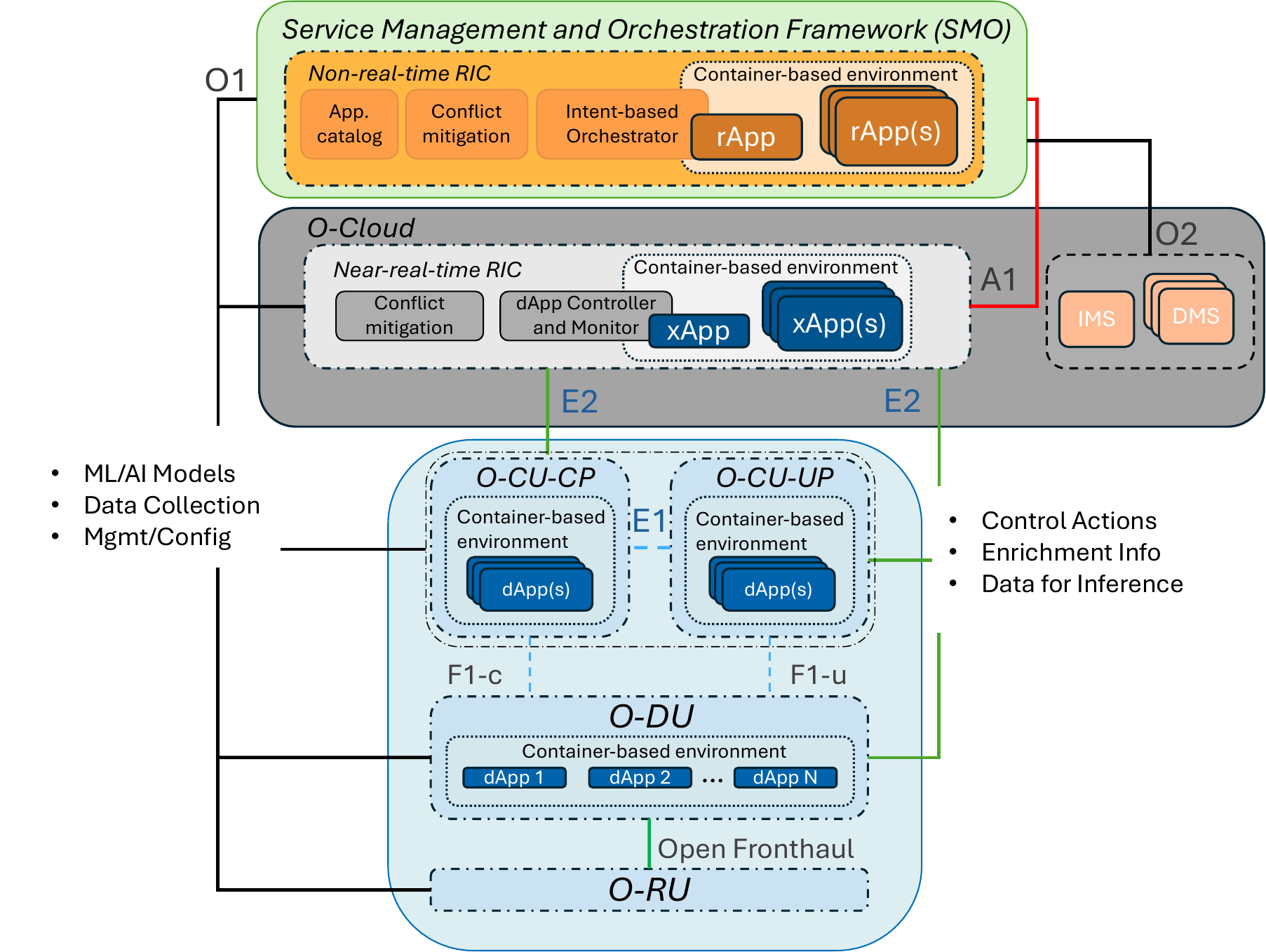}
\caption{\gls{ORAN} architecture.}
\label{fig:oranScheme}
\end{figure*}

To implement an effective frequency-affinity scheduling policy, accurate runtime monitoring of critical computational parameters is essential. We therefore developed a real-time monitoring tool using Linux’s \texttt{perf} subsystem, capable of extracting detailed thread-level execution metrics from \gls{CPU} hardware counters. Drawing from the insights presented in~\cite{AIRIC}, we identified the following performance metrics as critical indicators for assessing \gls{RAN} thread behavior and energy efficiency:

\begin{itemize}
    \item \textbf{\gls{CPU} utilization:} The percentage of time a thread actively runs on a processor.
    \item \textbf{Context switches:} The frequency at which threads are interrupted and resumed, negatively affecting performance due to overhead.
    \item \textbf{Instructions Per Cycle (IPC):} Efficiency measure indicating how effectively the \gls{CPU} executes instructions independently of frequency.
    \item \textbf{Misses Per Kilo Instruction (MPKI):} The frequency of cache misses per thousand executed instructions, representing memory inefficiencies that negatively impact execution latency and power consumption.
\end{itemize}

Based on this context, we clearly define our control problem as follows:

\vspace{0.5em}
\noindent\textbf{Problem Statement:}
We formulate the problem as an online constrained optimization targeting a subset of CPU cores dedicated to softwarized RAN workloads. The objective is to minimize dynamic power consumption while ensuring that all real-time processing deadlines are met and end-to-end throughput remains within acceptable deviation.

\vspace{0.5em}
\noindent\textit{Given}:
\begin{itemize}
    \item A set of CPU cores $\mathcal{C} = \{c_1, c_2, \dots, c_N\}$ allocated to the O-DU.
    \item A set of RAN processing threads $\mathcal{T} = \{\tau_1, \tau_2, \dots, \tau_M\}$.
    \item Real-time telemetry from Linux performance counters: CPU utilization, IPC, MPKI, context switches.
\end{itemize}

\vspace{0.5em}
\noindent\textit{Control Variables}:
\begin{itemize}
    \item $\pi: \mathcal{T} \rightarrow \mathcal{C}$ \quad (thread-to-core affinity mapping)
    \item $f_i \in [f_{\min}, f_{\max}]$ \quad (operating frequency per core $c_i$)
    \item $I_i \in \{0, 1\}$ \quad (core isolation indicator: 1 if $c_i$ is isolated)
\end{itemize}

\vspace{0.5em}
\noindent\textit{Objective Function}:
\begin{equation}
\min_{\pi, \{f_i\}, \{I_i\}} \quad E = \sum_{i=1}^{N} P(f_i) \cdot u_i
\end{equation}

where $P(f_i) = P_s + k f_i^2$ is the power model of core $c_i$, and $u_i$ is the utilization of $c_i$.

\vspace{0.5em}
\noindent\textit{Constraints}:
\begin{align}
    \forall \tau_j \in \mathcal{T}: &\quad \text{Latency}(\tau_j) \leq T_{\text{TTI}} \\
    &\quad |\text{Throughput}_{\text{measured}} - \text{Throughput}_{\text{baseline}}| \leq \delta \\
    &\quad \pi(\tau_j) \in \mathcal{C}_{\text{active}} \quad \text{(respect isolation: } I_i = 0)
\end{align}

\vspace{0.5em}
\noindent
To operationalize this optimization in real-time environments, we implement the control logic as a lightweight, containerized application co-located with the O-DU execution environment. The following section details the architecture, deployment model, and execution flow of the proposed dApp, highlighting the mechanisms that ensure minimal overhead and compliance with O-RAN specifications.


\section{Proposed System}
\label{sec:architecture}

The \gls{ORAN} architecture decomposes the \gls{RAN} into \glspl{RU}, \glspl{DU}, and \glspl{CU} interconnected through open interfaces (Figure~\ref{fig:oranScheme}). This modularisation enables vendor-agnostic deployments, yet shifts time-critical baseband execution to commodity CPUs inside the \gls{DU}, where static, worst-case provisioning is common and energy-inefficient under variable traffic.

Existing energy-saving approaches in the state of the art typically operate as host-level tweaks (e.g., governors, kernel patches) that are effective locally but remain outside a standardised management framework. As a result, they cannot be orchestrated, audited, or coordinated with radio policies and service objectives. O-RAN provides the missing integration layer; however, \glspl{rApp} (non-RT RIC) and \glspl{xApp} (near-RT RIC) act at seconds and tens-of-milliseconds timescales, respectively, which is insufficient for slot-level CPU actuation within the \gls{TTI} budget.

To close this gap, recent O-RAN specifications introduce \emph{distributed applications} (dApps), lightweight components executed on the O-Cloud node hosting the \gls{DU} and connected to the radio stack via the E3 interface. dApps access OS-level telemetry with microsecond granularity and can apply CPU-level actions (affinity, frequency, isolation) within the per-slot deadline, while remaining visible to the O-RAN management plane. Crucially, their state can be summarised upstream and aligned with longer-horizon objectives from \glspl{xApp} and \glspl{rApp}, enabling cross-timeframe optimisation rather than isolated host-side control.

Guided by the optimisation in Section~\ref{sec:problem}, the design follows three principles: (i) \emph{locality}: actuation co-located with \gls{ODU} threads to avoid E2 latency; (ii) \emph{vendor-agnosticism}: exclusive use of user-space knobs and standard telemetry (e.g., \texttt{perf}); and (iii) \emph{composability}: export of aggregated CPU state for coordination with higher-layer controllers.

\begin{figure}[tb]
\centering
\includegraphics[scale=0.45]{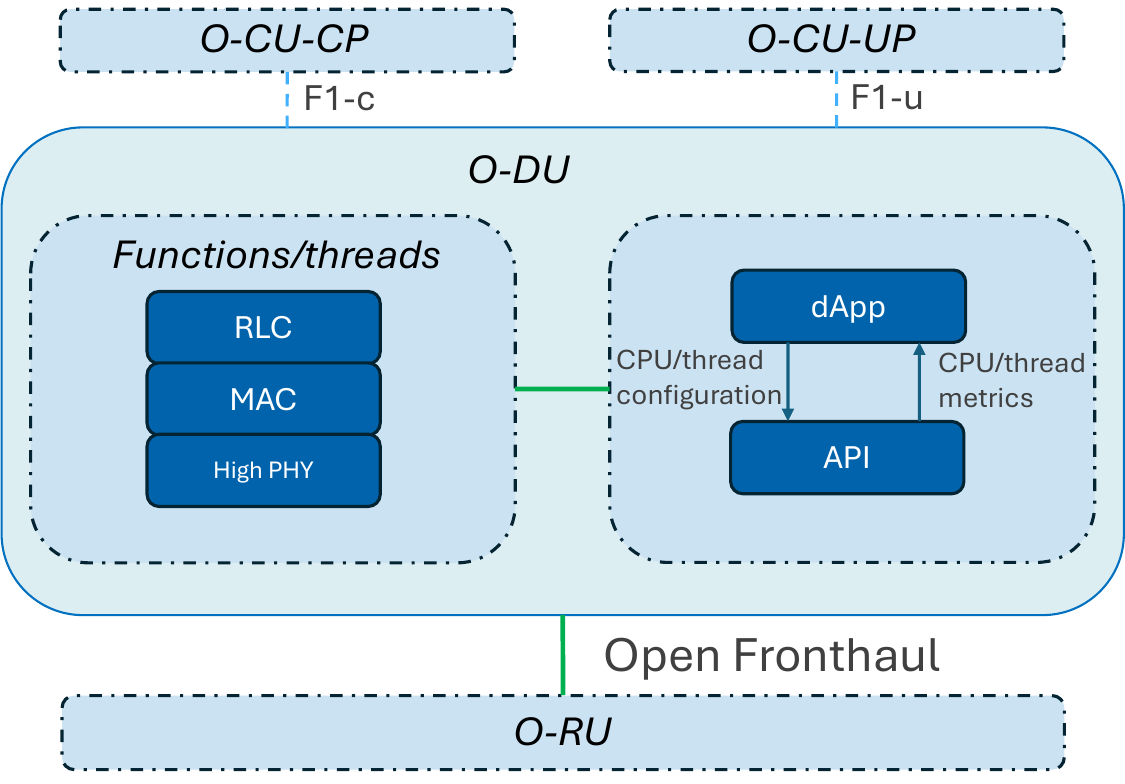}
\caption{Proposed dApp deployment architecture.}
\label{fig:dApp}
\end{figure}

Figure~\ref{fig:dApp} focuses on the on-node deployment.  
A \emph{telemetry container} gathers hardware counters and scheduler statistics, exposing a local API that a \emph{control container} (the dApp) polls to evaluate the constraints from Section~\ref{sec:problem}.  
The dApp then applies thread–core affinity, governor overrides, and core-isolation flags via user-space interfaces, requiring neither kernel modifications nor changes to the RAN stack, and adding only negligible overhead.

The life-cycle anchoring of these components is handled at system level by the near-RT RIC blocks shown in Figure~\ref{fig:oranScheme}: the \textit{dApp Controller \& Monitor} registers instances, distributes policies, and supervises health, whereas the \textit{Conflict Mitigation} xApp arbitrates CPU-level intents against concurrent radio objectives issued by other xApps.  
This mediation aligns sub-TTI CPU actions with the tens-of-milliseconds control loops of xApps and the longer-horizon policies of \glspl{rApp}, enabling cross-time-frame optimisation while preventing policy clashes.

The dApp operates independently of RAN vendor software, relying solely on OS-level telemetry (e.g., performance counters, scheduler statistics) and user-space control knobs (e.g., CPU affinity and governor tuning). This approach enables direct CPU-level actuation without modifying RAN stack internals or kernel behavior.

In the current implementation, control decisions are derived from a rule-based heuristic designed to balance energy savings and latency constraints. The controller dynamically reallocates threads across CPU cores and adjusts frequency scaling policies based on observed processing demand, with changes enforced at runtime. While the optimization formulation in Section~\ref{sec:problem} allows for more advanced control algorithms, the current heuristic approach demonstrates feasibility with minimal overhead.

\section{Experimental Evaluation}
\label{sec:evaluation}

Here, the \gls{dApp} will work with two different real-world 5G softwarized implementations implementing a frequency control strategy under varying traffic and computational loads.

To validate the proposed approach, experiments were conducted on a testbed implementing the \gls{ORAN} Split 8 architecture, where the \gls{DU} carries out both the high and low \gls{PHY} and the \gls{RU} is limited to the \gls{RF} chain. Thus, this split particularly relevant for enabling real-time control and computational offloading. In regard to the testbed, the most relevant parameters for the experiments are summarized in Table~\ref{tab:tablaParametros}.

The test environment is based on the open-source srsRAN Project, which is a complete 5G \gls{RAN} solution, featuring \gls{OCU} and \gls{ODU} aligning with 3GPP release 17, supporting FDD/TDD and all FR1 bands in all bandwidths \cite{srsranDocs}. The srsRAN \gls{RAN} stack is run on a computer with ArchLinux as \gls{OS} running the Linux Kernel 6.10.2, equipped with a $Intel(R)\ Core^{TM}\ i9-14900K$ \gls{CPU} with eight cores with \gls{SMT} disabled. Furthermore, for the use of Split 8, the USRP N300 from Ettus \cite{ettus_usrp_n300} has been used as \gls{SDR}. This \gls{SDR} that contains the \gls{RF} frontend, \gls{DAC} and \gls{ADC} to process digital samples. As for the \gls{UE}, a laptop with ArchLinux has been used, connecting to the network through a SIMCOM SIM8380G-M2 modem \cite{SIMCom}. 

Moreover, \gls{TCP} throughput and latency measurements have been conducted using \textit{iperf} and \textit{ping}, respectively. Nevertheless, latency tests have not shown significant results in the experiments carried out, in line with the results from \cite{Perfprofiling}, since the air interface might masks subtle differences in latency.

\begin{table}[t]
    \centering
    \vspace{-5pt}
    \caption{Considered network parameters}
    \label{tab:tablaParametros}
    \begin{tabular}{ m{0.35\columnwidth} >{\centering\arraybackslash}p{0.35\columnwidth} }
        \toprule
        \textbf{Parameter}               & \textbf{Value}      \\
        \midrule
        \rowcolor{verylightgray}\textbf{BW}           & 50 MHz     \\
        \textbf{Band} & 78           \\
        \rowcolor{verylightgray}\textbf{\#Tx Antennas}         & 1   \\
        \textbf{\#Rx Antennas}         & 1   \\
        \rowcolor{verylightgray}\textbf{Tx\ Gain}      & 65 \\
        \textbf{Rx\ Gain}     & 45     \\
        \rowcolor{verylightgray}\textbf{Available Cores}    & 8  \\
        \textbf{\#gnbs}    & [1-5]  \\
        \rowcolor{verylightgray}\textbf{Mux\ Type}    &  TDD \\
        \bottomrule
        
    \end{tabular}
    \vspace{-5pt}
\end{table}

\begin{figure}[b]
\centering
\includegraphics[width=2in]{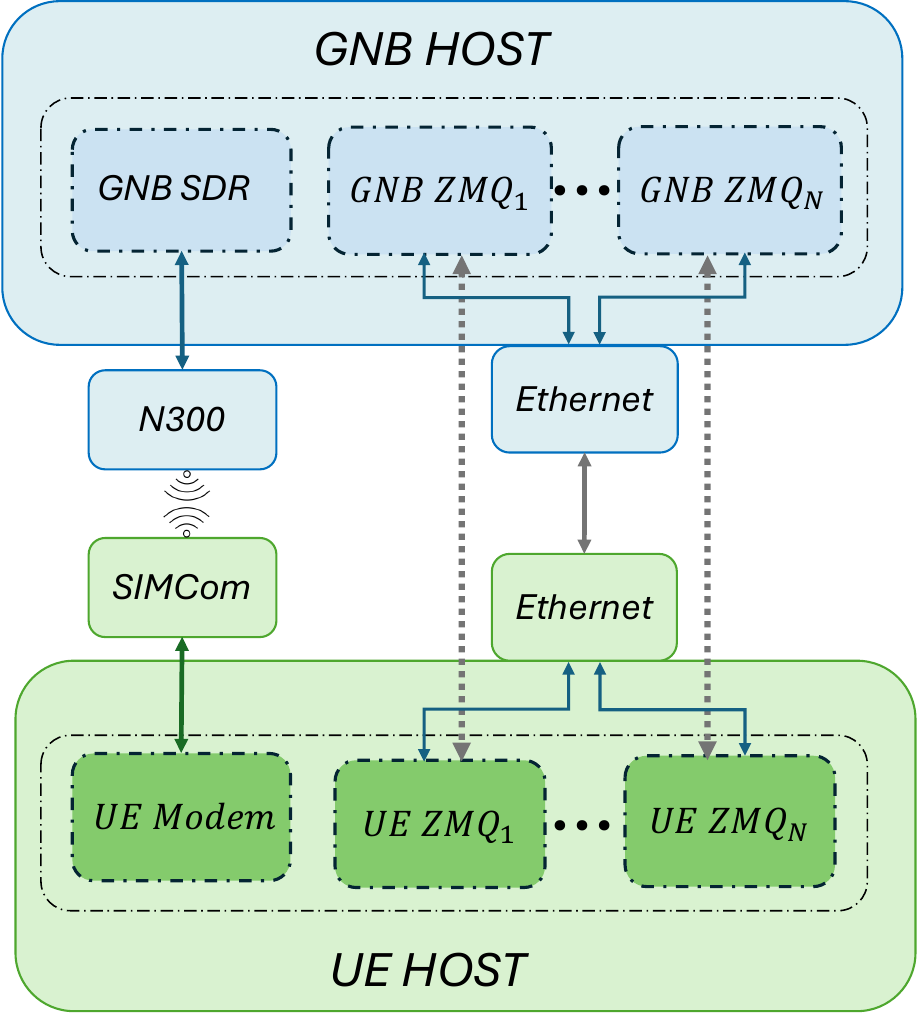}
\caption{Deployment for evaluating Noisy Neighbor effects.}
\label{fig:srsTestBed}
\end{figure}
\subsection{Noisy-Neighbour Scenario}

\begin{figure*}[h]
    \centering
    \begin{subfigure}[b]{0.4\textwidth}
        \centering
        \includegraphics[width=\linewidth]{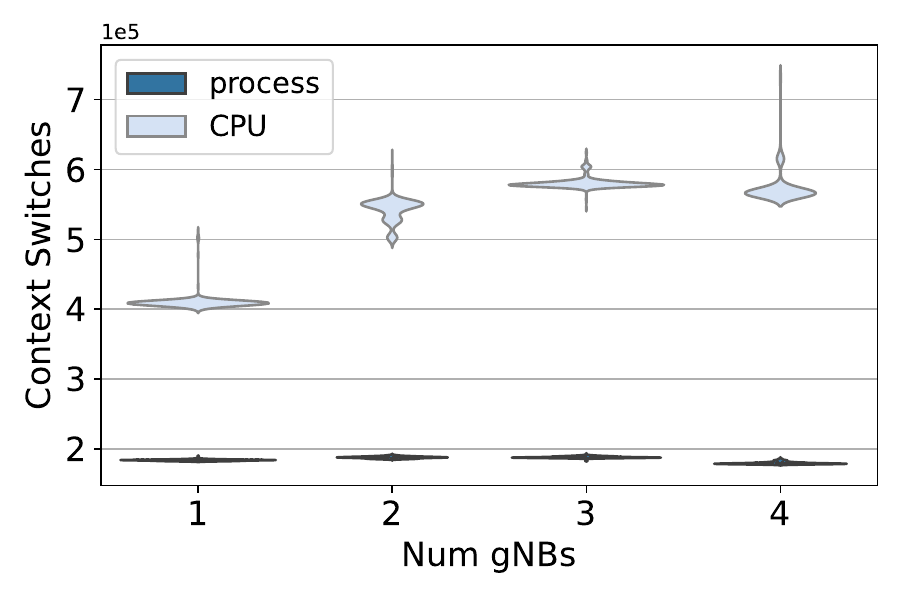}
        \caption{Context switches}
        \label{fig:srs_CS_NN}
    \end{subfigure}
    \hspace{0pt}
    \begin{subfigure}[b]{0.4\textwidth}
        \centering
        \includegraphics[width=\linewidth]{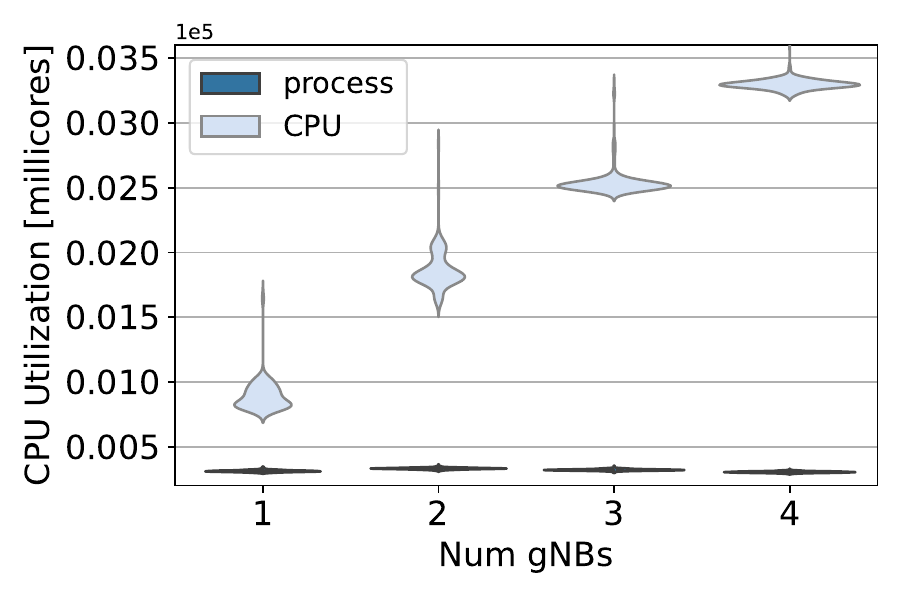}
        \caption{CPU utilization}
        \label{fig:srs_CPU_NN}
    \end{subfigure}

    \begin{subfigure}[b]{0.4\textwidth}
        \centering
        \includegraphics[width=\linewidth]{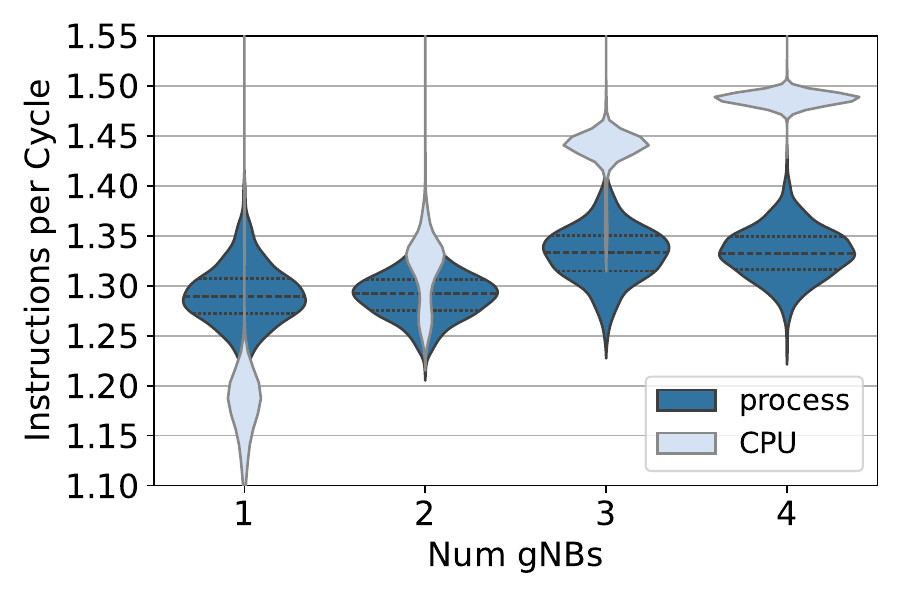}
        \caption{Instructions per Cycle}
        \label{fig:srs_IPC_NN}
    \end{subfigure}
    \hspace{0pt}
    \begin{subfigure}[b]{0.4\textwidth}
        \centering
        \includegraphics[width=\linewidth]{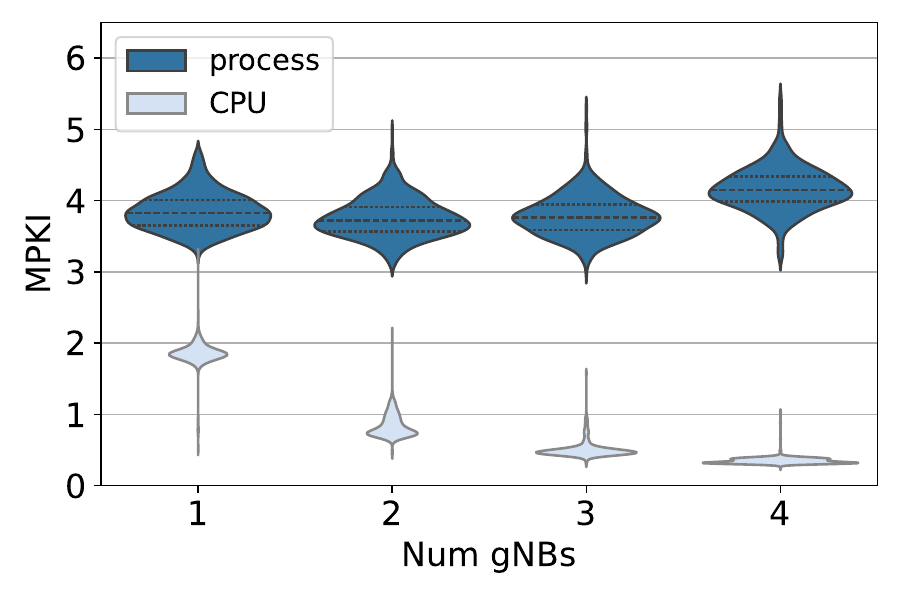}
        \caption{Misses per 1000 instructions}
        \label{fig:srs_MPKI_NN}
    \end{subfigure}

    \caption{Metrics accumulated by the process and accumulated by the CPU pool in the srsRAN deployment.}
    \label{fig:NoisyNeighbourInmunity}
\end{figure*}

Co-locating several containerised \glspl{gNB} on the same host is common in practical deployments and can trigger the \emph{noisy-neighbour} effect, where a burst of activity in one instance perturbs the real-time behaviour of its peers.  To quantify this impact, up to five independent srsRAN \gls{gNB} containers were launched on the eight physical cores reserved for the O-DU.  One container, the \emph{foreground} cell, used an over-the-air USRP N300 and a SIMCOM UE modem, while the remaining cells generated background traffic through ZeroMQ radios, reproducing contention without extra RF hardware (Figure~\ref{fig:srsTestBed}).  Amarisoft could not be included because the available licence supports only a single \gls{gNB}.

Figure~\ref{fig:NoisyNeighbourInmunity} tracks four CPU-level metrics as the number of background \glspl{gNB} increases.  Although srsRAN’s default affinity keeps process-level counters stable, core-level measurements reveal clear contention: context switches climb rapidly until three \glspl{gNB} and then plateau; utilisation grows almost linearly; meanwhile, higher IPC and lower MPKI suggest that additional threads exhibit a more cache-friendly access pattern.  This controlled stress test provides the baseline against which the proposed dApp’s affinity and frequency policies are evaluated in Section~\ref{sec:results}.


In summary, the results from the experimentation with Noisy Neighbors suggest that srsRAN offers a thread affinity default strategy that enables it to run multiple \glspl{gNB} without degrading process-level performance. Furthermore, \gls{CPU}-level metrics indicated that computational parameters stabilizes with a few \glspl{gNB}, with the exception of \gls{CPU} utilization which increases linearly. In addition, it can be seen that \gls{IPC} increases gradually with \gls{CPU} utilization as the \gls{CPU} maximizes its internal utilization, increasing instructions output per cycle. Lastly, in contrast to what is observed in \cite{AIRIC} where the number of \gls{OAI} \glspl{gNB} increase \gls{CPU} utilization exponentially, the linear growth in srsRAN might be indicative that the affinity approach used by the latter is more optimal for deploying multiple \glspl{gNB}.

\subsection{Energy-Saving dApp Results}\label{sec:results}

\begin{figure*}[t!]
\centering
\includegraphics[width=4in]{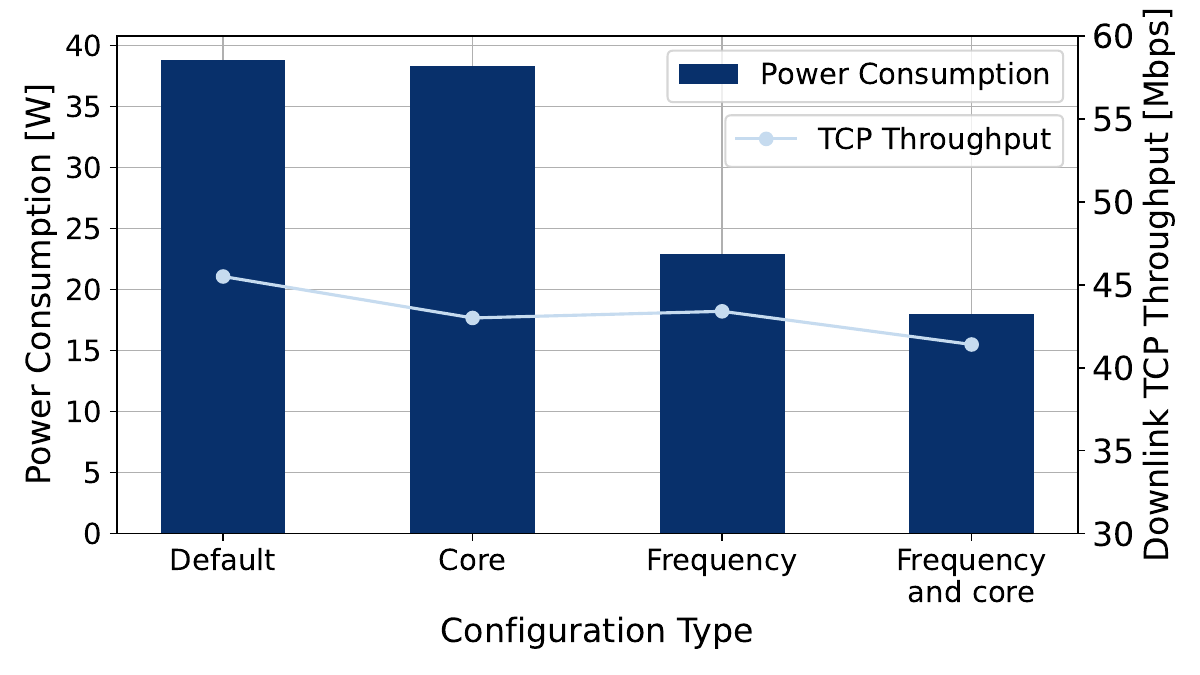}
\caption{Power consumption using core affinity and frequency control strategies on srsRAN 5G.}
\label{fig:energyAffinity}
\end{figure*}

Figure~\ref{fig:energyAffinity} compares four execution policies applied to the same srsRAN gNB container while maintaining identical traffic and RF conditions.  
\emph{Configuration I}, the unmodified system with the Linux \texttt{performance} governor, keeps all cores at peak frequency, draws \SI{38}{W}, and delivers \SI{50}{Mbps}.  The high power figure reflects the quadratic term of the dynamic component in~\eqref{eq:powerEq}; moreover, the absence of affinity inflates context-switch and MPKI counters (Figures \ref{fig:srs_CS_NN}–\ref{fig:srs_MPKI_NN}), evidencing cache thrashing as threads migrate across cores.

\emph{Configuration II} introduces static affinity while retaining the \texttt{performance} governor.  Thread locality reduces LLC conflicts, yet pinned cores remain locked at maximum frequency even during low-demand slots, so power only drops to \SI{33}{W}.  Throughput falls to \SI{47}{Mbps} because the scheduler no longer redistributes slack to absorb jitter from background processes, confirming that isolation by itself does not guarantee efficiency.

\emph{Configuration III} relaxes affinity but enables on-demand frequency scaling.  Average consumption declines sharply to \SI{26}{W}; the governor exploits burst slack to enter lower P-states, and throughput remains almost unchanged (\SI{49}{Mbps}).  Nevertheless, the lack of pinning causes occasional imbalance among cores, raising variance in per-core utilisation and slightly increasing tail latency (not shown in the figure).

\emph{Configuration IV} combines affinity with frequency scaling.  By clustering cache-intensive threads and allowing the governor to follow demand, dynamic power is minimised without sacrificing locality.  Average draw reaches the floor at \SI{19.5}{W}, a \SI{49}{\percent} reduction from the baseline, while throughput improves marginally to \SI{51}{Mbps}.  Context-switch counts drop below \(1.2\times10^{5}\), IPC stabilises above 1.5, and MPKI halves compared with Configuration I, demonstrating that the joint policy eliminates most scheduler noise.

Two observations reinforce the suitability of the proposed approach.  First, all measurements were taken with the dApp running in user space and communicating via its local API, adding less than \SI{0.3}{\percent} CPU overhead; hence, the gains originate from better scheduling rather than measurement artifacts.  Second, slot-level latency never exceeded the \SI{1}{ms} TTI budget in any configuration, confirming that frequency transitions and affinity updates inserted by the dApp do not compromise real-time deadlines.  The results therefore validate the premise that fine-grained, OS-driven control, when coordinated through the O-RAN dApp framework, can achieve substantial energy savings while maintaining or even enhancing user-plane performance.

\section{Conclusions and Future Work}
\label{sec:conclusion}

This work introduced an \gls{ORAN}-aligned \gls{dApp} that closes the control-loop gap between host-level \gls{CPU} management and RAN-wide orchestration. The proposed approach introduces a \gls{dApp} deployed directly at the E2 node (specifically, the \gls{ODU}), capable of observing and reacting to fine-grained \gls{OS}-level telemetry without modifying the \gls{RAN} software stack. By collecting metrics such as context switches, \gls{IPC}, and \gls{MPKI} through the Linux's \textit{perf} tool, the \gls{dApp} dynamically adjusts the \gls{CPU} operating frequency and assigned cores in response to workload conditions. This strategy is designed to remain agnostic to the underlying \gls{RAN} implementation and operates fully within the timing constraints of \gls{ODU}-level processing threads.

The experimental results on a srsRAN deployment demonstrate measurable savings in power consumption without compromising real-time execution performance. The analysis also shows that \gls{CPU} inefficiencies such as excessive thread migrations or memory contention correlate strongly with energy waste, highlighting the potential benefits of future affinity and isolation strategies. While these were not directly applied in this work, their effects were characterized through extensive profiling.

Several directions for future work arise from the findings of this study. First, expanding the set of observable metrics by incorporating additional telemetry such as memory bandwidth saturation, last-level cache contention, and \gls{NUMA} locality. These inputs could enable finer-grained classification of \gls{CPU} states and task behaviors, serving as a foundation for adaptive scheduling decisions. Secondly, analyzing thread-level metrics and exploring clustering techniques to group threads with similar execution signatures. By applying thread-level clustering over temporal and structural metrics, it becomes possible to assign affinity configurations more efficiently, without evaluating the full set of computational performance counters on every scheduling interval. This dimensionality reduction can significantly lower the computational overhead of real-time decision making, while improving the isolation of critical tasks. Lastly, investigating fairness at the \gls{OS}-level scheduling objective in the context of shared \gls{RAN} environments. Here, characterizing fairness not only as a constraint, but as a measurement of imbalance or degradation, may help identify contention phases and trigger corrective actions which could lead to adaptive scheduling strategies that better reflect the service-level priorities of co-located cloudified base stations.

Together, these extensions aim to strengthen the \gls{dApp}'s ability to perform scalable, interpretable, and energy-efficient control of \gls{CPU} resources in line with \gls{ORAN} deployment principles. By continuing to exploit fine-grained telemetry at the node level, the objective also aims to enhance orchestration capabilities without increasing system complexity or compromising interoperability.

\bibliography{literature/bibliography}

\newpage
\begin{IEEEbiographynophoto}{Francisco Crespo}
 received his degree in Telecommunication Technologies Engineering at the University of Malaga, Spain, in 2022. He works as a researcher at the University of M\'alaga, focusing on the management of computational resources in shared virtualised environments.
\end{IEEEbiographynophoto}

\begin{IEEEbiographynophoto}{Javier Villegas}
received his degree in Telecommunications Systems Engineering and his M.Sc. degrees in telecommunication engineering and in telematic engineering from the University of M\'alaga, Spain. Currently, he is working as a Assistant Professor with the Department of Communications Engineering at the University of M\'alaga, where he is pursuing a Ph.D.
\end{IEEEbiographynophoto}

\begin{IEEEbiographynophoto}{Carlos Baena}
obtained his Ph.D. in Mobile Networks from the University of M\'alaga, Spain. His research specializes in the optimization of end-to-end (E2E) network performance through service-based approaches. He focuses on enhancing the user experience , particularly in applications related to video streaming and gaming, by leveraging machine learning (ML) techniques and key quality indicators (KQI) to optimize network resource management and overall performance.
\end{IEEEbiographynophoto}

\begin{IEEEbiographynophoto}{Eduardo Baena}
is a Postdoctoral Research Fellow at Northeastern University. He holds a Ph.D. in Telecommunication Engineering from the University of M\'alaga (UMA), where he also served as a lecturer and researcher. Between 2010 and 2017, he worked in various technical and leadership roles in the international private sector. At UMA, he contributed to several H2020 research projects and served as Co-PI on multiple nationally and regionally funded initiatives. His current research focuses on AI-driven cellular networks, 5G/6G architectures, O-RAN, NTNs, and LEO-based edge computing.
\end{IEEEbiographynophoto}

\begin{IEEEbiographynophoto}{Sergio Fortes}
is Associate Professor at the University of M\'alaga, from which it holds a M.Sc. (2010) and a Ph.D. (2017) in Telecommunication Engineering. He began his career being part of main european space agencies (DLR, CNES, ESA) and Avanti Communications plc, where he participated in various research and consultant activities on broadband and aeronautical satellite communications. In 2012, he joined the University of M\'alaga, where his topics of interest include cellular communications, satellite systems, smart-city, self-organizing / zero-touch networks (SON/ZSN), cloud robotics, and advanced applications of AI and machine learning techniques.
\end{IEEEbiographynophoto}

\begin{IEEEbiographynophoto}{Raquel Barco}
holds a M.Sc. and a Ph.D. in Telecommunication Engineering
from the University of M\'alaga. From 1997 to 2000, she worked at Telef\'onica in Madrid (Spain) and at the European Space Agency (ESA) in Darmstadt (Germany). In 2000, she joined the University of M\'alaga, where she is currently Full Professor. She took part as researcher in a Nokia Competence Center on Mobile Communications for three years. She has led projects with the main mobile communications operators and vendors for a value\textgreater 15 million €, she is author of 7 patents and has published more than 150 high impact journals and conferences.
\end{IEEEbiographynophoto}


\end{document}